\documentclass[article,prl,balancelastpage,twocolumn,showpacs]{revtex4}
\usepackage{makeidx}
\usepackage{amssymb}
\usepackage{dcolumn}
\usepackage{graphicx}
\usepackage{acronym}

\input{tcilatex}

\begin{document}

\title{Localization of fremions in rotating electromagnetic fields}
\author{B. V. Gisin }
\affiliation{IPO, Ha-Tannaim St. 9, Tel-Aviv 69209, Israel. E-mail: gisin@eng.tau.ac.il}
\date{\today }

\begin{abstract}
\noindent Parameters of localization are defined in the lab and rotating
frame for solutions of the Dirac equation in the field of a traveling
circularly polarized electromagnetic wave and constant magnetic field. The
radius of localization is of the order of the electromagnetic wavelength and
lesser.
\end{abstract}

\pacs{03.65.Pm, 03.65.Ta, 31.30.jx, 06.20.Jr\hspace{20cm}}
\maketitle

\section{Introduction}

Recently 3D transformation for rotating frames was deduced using the
'invariance of the Dirac equation under rotation with a constant frequency
provided that the cylindrical radius is invariable' \cite{BVG}. The
transformation was obtained from very general assumptions and may be used in
variety applications. With help of the transformation the wave function of \
Dirac equation can be translated to a rotating frame.

In the paper we study localization of fermions on an example of exact
solutions of Dirac's equation in the field of a traveling circularly
polarized electromagnetic wave and constant magnetic field.\ 

We start from a brief description of the transformation and exact localized
solutions of the Dirac equation.\ \ \ \ \ \ \ \ \ \ \ \ \ \ \ \ \ \ \ \ \ \
\ \ \ \ \ \ \ \ \ \ \ \ \ \ \ \ \ \ \ \ \ \ \ \ \ \ \ \ \ \ \ \ \ \ \ \ \ \
\ \ \ \ \ \ \ \ \ \ \ \ \ \ \ \ \ \ \ \ \ \ \ \ \ \ \ 

\section{The transformation}

We use the dimensionless units 
\begin{equation}
t\rightarrow \frac{ct}{\lambdabar },\text{ \ }(x,y,z)\rightarrow \frac{%
(x,y,z)}{\lambdabar },  \label{nc}
\end{equation}%
$\lambdabar $ is the Compton wavelength $\lambdabar =\hbar /mc$.

The transformation of the cylindrical coordinates $\varphi ,z,t$ has the
form \cite{BVG}%
\begin{eqnarray}
\tilde{\varphi} &=&\varphi +z\Omega -\Omega t  \label{trp} \\
\tilde{z} &=&\frac{-r^{2}\Omega \varphi }{\sqrt{1-r^{2}\Omega ^{2}}}+z\sqrt{%
1-r^{2}\Omega ^{2}}+\frac{r^{2}\Omega ^{2}t}{\sqrt{1-r^{2}\Omega ^{2}}},
\label{tr} \\
\tilde{t} &=&\frac{-r^{2}\Omega \varphi }{\sqrt{1-r^{2}\Omega ^{2}}}+\frac{t%
}{\sqrt{1-r^{2}\Omega ^{2}}}  \label{trt}
\end{eqnarray}%
where the normalized frequency $\Omega \rightarrow \Omega \lambdabar /c.$ 
\begin{eqnarray}
\cosh \Phi &=&\frac{1}{\sqrt{1-r^{2}\Omega ^{2}}},\ \sinh \Phi =\frac{%
r\Omega }{\sqrt{1-r^{2}\Omega ^{2}}}  \label{chsh} \\
\sin \Phi _{1} &=&-r\Omega ,\text{ \ }\cos \Phi _{1}=\sqrt{1-r^{2}\Omega ^{2}%
}.
\end{eqnarray}%
The determinant of this transformation equals $1.$

This 3D transformation is one-parametrical, that is variables with and
without the tilde coincide in absence of rotation.

A distinguish feature of Eqs. (\ref{tr}) is an upper boundary for radius $%
r^{2}\Omega ^{2}\leq 1.$ In the non-normalized units this inequality is
equivalent to 
\begin{equation}
r\leq \frac{\lambda }{2\pi },  \label{rlim}
\end{equation}%
where $\lambda $ is the wavelength corresponding to the frequency $\Omega $.

The invariant transformation of the Dirac equation is realized by
multiplication of this equation by operator%
\begin{equation}
P=\exp (\frac{1}{2}\alpha _{2}\alpha _{3}\Phi _{1}+\frac{1}{2}\alpha
_{2}\Phi ).
\end{equation}%
For the transformation of spinor the operator $\tilde{P}=\beta P\beta $ is
used, where $\alpha _{2},$ $\alpha _{3},$ $\beta $ are Dirac' matrices \cite%
{BVG}.

\section{Solutions of Dirac' equation in rotation electromagnetic field}

Consider Dirac's equation%
\begin{equation}
\{-i\frac{\partial }{\partial t}-i\mathbf{\alpha }\frac{\partial }{\partial 
\mathbf{x}}-\mathbf{\alpha \mathbf{A}}+\beta \}\Psi =0.  \label{Dir}
\end{equation}%
in the electromagnetic field with the potential 
\begin{eqnarray}
A_{x} &=&-\frac{1}{2}H_{z}y+\frac{1}{\Omega }H\cos (\Omega t-kz),  \label{Ax}
\\
A_{y} &=&\frac{1}{2}H_{z}x+\frac{1}{\Omega }H\sin (\Omega t-kz).  \label{Ay}
\end{eqnarray}%
This potential describes a traveling circularly polarized electromagnetic
wave of the frequency $\Omega $ propagating along constant magnetic field $%
H_{z}$. In the normalized dimensionless units the propagation constant $%
k=\Omega .$ The normalized potential is defined as%
\begin{equation}
\mathbf{A\rightarrow }\frac{e\lambdabar }{c\hbar }\mathbf{A,}\text{ \ }%
\mathbf{H\rightarrow }\frac{e\lambdabar ^{2}}{c\hbar }\mathbf{H.}  \label{A}
\end{equation}

The Dirac's equation (\ref{Dir}) has exact solutions localized in the cross
section perpendicular to the propagation direction of the wave \cite{BG}. In
the lab frame only non-stationary states are possible. In contrast to that
stationary states exist in a rotating frame.

The solutions in the lab frame can be presented as follows%
\begin{equation}
\Psi =\exp [-iEt+ipz-\frac{1}{2}\alpha _{1}\alpha _{2}(\Omega t-\Omega
z)+D]\psi ,  \label{sol0}
\end{equation}%
\begin{eqnarray}
D &=&-\frac{d}{2}r^{2}-id_{2}\tilde{x}+d_{2}\tilde{y}, \\
d &=&-\frac{1}{2}H_{z},\text{ \ }d_{2}=\frac{\mathcal{E}_{0}h}{2(\mathcal{E}-%
\mathcal{E}_{0})}.\text{ \ }
\end{eqnarray}%
In the normalized units $d_{2}\rightarrow d_{2}\lambdabar ,$ $d\rightarrow
d\lambdabar ^{2}$ $E\rightarrow E/mc^{2},$ $p\rightarrow p/mc$ . $E$ obeys
the characteristic equation%
\begin{equation}
\mathcal{E}(\mathcal{E}+2p-\Omega )-1-\frac{\mathcal{E}h^{2}}{\mathcal{E}-%
\mathcal{E}_{0}}=0,  \label{Ceq}
\end{equation}%
\begin{equation}
\mathcal{E}=E-p,\text{ }\mathcal{E}_{0}=\frac{2d}{\Omega },\text{\ \ }h=%
\frac{1}{\Omega }H,  \label{E0}
\end{equation}%
the parameter $h$ usually is small.

The parameter $\mathcal{E}_{0}$ in the non-normalized units is defined as%
\begin{equation}
\mathcal{E}_{0}=-\frac{2\mu H_{z}}{\hbar \Omega },  \label{EG}
\end{equation}%
where $\mu =e\hbar /(2mc)$ is the Bohr magneton. Equating $2/\mathcal{E}_{0}$
to g-factor turns the definition (\ref{EG}) in the classical condition of
the magnetic resonance. The charge, for definiteness, is assumed to be
negative $e=-|e|$. The equality (\ref{EG}) is kept by the sign change of $e,$
provided the constant magnetic field $H_{z}$ has the opposite direction or
by the the opposite polarization of the electromagnetic wave$.$

Below, for definiteness, we assume that $\mathcal{E}_{0}$ is always positive.

The spinor $\psi $ depend of $\tilde{x},\tilde{y}$. A constant spinor
describe the ground state, a spinor polynomial corresponds to excited state.

The spinor $\psi $ corresponding to the ground state has shape%
\begin{equation}
\psi =N\left( 
\begin{array}{c}
h\mathcal{E} \\ 
-(\mathcal{E}+1)(\mathcal{E}-\mathcal{E}_{0}) \\ 
h\mathcal{E} \\ 
-(\mathcal{E}-1)(\mathcal{E}-\mathcal{E}_{0})%
\end{array}%
\right) ,  \label{solphi}
\end{equation}%
$N$ is the normalization constant defined by the normalization integral $%
\int \Psi ^{\ast }\Psi ds=1.$%
\begin{equation}
\text{ }N^{2}[h^{2}\mathcal{E}^{2}+(\mathcal{E}^{2}+1)(\mathcal{E}-\mathcal{E%
}_{0})^{2}]\frac{\pi }{d}\exp (\frac{d_{2}^{2}}{d})=1  \label{N}
\end{equation}%
We restrict ourselves the consideration of the ground state.

Spinor (\ref{solphi}) as well as spinors of excited states never can be
presented in the form a small and large two-component spinors. It means that
the spinor describes only relativistic fermion.

Typically $\mathcal{E}$\ is expanded in power series in $h^{2}.$ However,
there 'singular solutions' are possible. For such solutions $\mathcal{E}$\
is expanded in power series in $h$ 
\begin{equation}
\mathcal{E}_{1,2}\mathcal{=E}_{0}+h\mathcal{E}_{1,2}+h^{2}\mathcal{E}%
_{2}+\ldots ,\text{ \ }\mathcal{E}_{1,2}=\pm \frac{\mathcal{E}_{0}}{\sqrt{%
\mathcal{E}_{0}^{2}+1}},  \label{Eh}
\end{equation}%
This expansion corresponds to a pair solutions. Odd terms in (\ref{Eh}) have
positive and negative signs for one and other solution in the pair. In the
first approximation%
\begin{equation}
d_{2}\approx \pm \frac{\sqrt{\mathcal{E}_{0}^{2}+1}}{2}.  \label{dpm}
\end{equation}

The necessary condition for existence of the expansion is a certain momentum
for both the states in the pair.%
\begin{equation}
p=\frac{1}{2}(\frac{1}{\mathcal{E}_{0}}-\mathcal{E}_{0})+\frac{1}{2}\Omega .
\label{p}
\end{equation}%
With this momentum the energy $E$ also coincide but with accuracy $\sim h$ 
\begin{equation}
E=\frac{1}{2}(\frac{1}{\mathcal{E}_{0}}+\mathcal{E}_{0})+\frac{1}{2}\Omega
+\ldots  \label{E}
\end{equation}

Eq. (\ref{Ceq}) is of third power. The third root of this equation is%
\begin{equation}
\mathcal{E=-}\frac{1}{\mathcal{E}_{0}}\mathcal{-}\frac{\mathcal{E}_{0}}{(1+%
\mathcal{E}_{0}^{2})}h^{2}+\ldots
\end{equation}

In the classical magnetic resonance, usually, a linearly oscillating
magnetic field is applied. Such a field is a combination of two circularly
polarized fields with opposite rotation. Fermion itself 'chooses' the
convenient polarization. The second polarization results in a weakly
dependence of spin on time, which may be neglected. But in assumed
experiments with the singular solutions the polarization should be taken
into account.

\section{Localization}

\subsection{Localization in the lab frame}

The parameter localization is defined from the integral 
\begin{equation}
\int_{-\infty }^{+\infty }\Psi ^{\ast }r^{2}\Psi dxdy
\end{equation}%
For the wave function (\ref{sol0}) this integral can be easily calculated.
In non-normalized units 
\begin{equation}
\sqrt{\overline{x^{2}}+\overline{y^{2}}}=\lambda \sqrt{\frac{\mathcal{E}%
_{0}^{2}+1}{\pi \mathcal{E}_{0}^{2}}}  \label{rlab}
\end{equation}

\subsection{Localization in the rotating frame}

As an example, we consider here the singular solutions. This case allows to
present parameter of localization in a simple form.

The parameter of localization in the rotating frame is defined by the ratio%
\begin{equation}
\frac{\int \tilde{\Psi}^{\ast }r^{2}\tilde{\Psi}rdrd\tilde{\varphi}}{\int 
\tilde{\Psi}^{\ast }\tilde{\Psi}rdrd\tilde{\varphi}}.  \label{rat}
\end{equation}

The procedure of the wave function definition in the rotating frame consist
of: multiplication of $\Psi $ by $\exp \frac{1}{2}\alpha _{1}\alpha
_{2}\varphi $ and operator $\tilde{P}$ by transition to the cylindrical
coordinates and the rotating frame. Further $\Psi $ should be multiplied by 
\begin{equation}
\exp (-\frac{1}{2}\alpha _{1}\alpha _{2}\tilde{\varphi})=\exp (-\tilde{%
\varphi}_{12})
\end{equation}%
for transition to the Cartesian coordinates in the rotating frame. As result
the wave function takes form%
\begin{equation}
\tilde{\Psi}=\exp [-\tilde{\varphi}_{12}+\frac{1}{2}\alpha _{2}\alpha
_{3}\Phi _{1}-\frac{1}{2}\alpha _{2}\Phi +\tilde{\varphi}_{12}+D]\psi 
\end{equation}

The limits of integration with respect to $\tilde{\varphi}$ in the rotating
frame are the same because of the condition (\ref{trp}), while the upper
limit for $r$ bounded by the condition (\ref{rlim}). After a simplification
the normalization integral $\int \tilde{\Psi}^{\ast }\tilde{\Psi}ds$ takes
the form%
\begin{equation}
2\pi \int_{r=0}^{1/\Omega }\psi ^{\ast }\frac{I_{0}(u)+\alpha _{1}r\Omega
I_{0,u}}{\sqrt{1-r^{2}\Omega ^{2}}}Y(-w)\psi rdr,  \label{N1}
\end{equation}%
where $I_{0}(u)$ is the modified Bessel functions of the first kind, $%
Y(w)=\exp (w)$; $u=2d_{2}r,$ $w=-dr^{2}$.

Equate%
\begin{equation}
r\Omega =\sin \theta ,
\end{equation}%
and use the parameter $\varkappa $%
\begin{equation}
\varkappa =\frac{c}{\Omega \lambdabar }=\frac{\lambda }{\lambda _{C}},\text{
\ }\lambda _{C}=2\pi \lambdabar .  \label{kap}
\end{equation}

For the singular solutions%
\begin{eqnarray}
\psi ^{\ast }\psi &=&4h^{2}\mathcal{E}_{0}^{2},\text{ \ }\psi ^{\ast }\alpha
_{1}\psi =\mp 4h^{2}\frac{\mathcal{E}_{0}^{3}}{\sqrt{1+\mathcal{E}_{0}^{2}}},
\\
u &=&\pm \varkappa \sqrt{\mathcal{E}_{0}^{2}+1}\sin \theta , \\
w &=&-\varkappa \frac{\mathcal{E}_{0}}{2}\sin ^{2}\theta .
\end{eqnarray}%
Denote%
\begin{eqnarray}
\eta &=&\int_{\theta =0}^{\pi /2}I_{0}Y\sin \theta d\theta , \\
\varsigma &=&\int_{\theta =0}^{\pi /2}I_{0,u}Y\sin ^{2}\theta d\theta , \\
\xi &=&\int_{\theta =0}^{\pi /2}I_{0}Y\sin \theta \cos ^{2}\theta d\theta .
\end{eqnarray}%
In order to carry out the calculation of ratio (\ref{rat}) the required
integrals in the numerator can be expressed as functions of parameters $\eta
,$ $\varsigma ,$ $\xi $%
\begin{eqnarray}
\int_{\theta =0}^{\pi /2}I_{0}Y\sin ^{3}\theta d\theta &=&\eta -\xi , \\
\int_{\theta =0}^{\pi /2}I_{0,u}Y\sin ^{4}\theta d\theta &=&\varsigma \mp 
\frac{\sqrt{\mathcal{E}_{0}^{2}+1}}{\mathcal{E}_{0}}\xi +\frac{1}{\varkappa 
\mathcal{E}_{0}}\varsigma ,
\end{eqnarray}%
These parameters cannot be calculated analytically. However, taken into
account that $|\varkappa |$ is very large $\sim 10^{9}$, this problem can be
solved by means of an asymptotic expansion of the parameters for large $%
|\varkappa |$.

Differentiate $\eta ,$ $\varsigma ,$ $\xi $ in respect to $\varkappa $
obtain after simplification the system of equation%
\begin{equation}
\varsigma _{,\varkappa }=\pm \sqrt{\mathcal{E}_{0}^{2}+1}\eta \mp \frac{1}{2}%
\sqrt{\mathcal{E}_{0}^{2}+1}\xi -\frac{\mathcal{E}_{0}}{2}\varsigma -\frac{3%
}{2\varkappa }\varsigma ,
\end{equation}%
\begin{equation}
\xi _{,\varkappa }=\frac{\mathcal{E}_{0}^{2}+1}{2\mathcal{E}_{0}}\xi \mp 
\frac{\sqrt{\mathcal{E}_{0}^{2}+1}}{2\varkappa \mathcal{E}_{0}}\varsigma -%
\frac{3}{2\varkappa }\xi +\frac{1}{2\varkappa }\eta ,
\end{equation}%
\begin{equation}
\eta _{,\varkappa }=\pm \sqrt{\mathcal{E}_{0}^{2}+1}\varsigma -\frac{%
\mathcal{E}_{0}}{2}\eta +\frac{\mathcal{E}_{0}}{2}\xi =0.
\end{equation}

Neglecting terms $\sim 1/\varkappa $ obtain the asymptotic expansion. With
accuracy $1/\varkappa $

\begin{eqnarray}
\xi &=&C_{1}\rho _{1}, \\
\eta &=&-\mathcal{E}_{0}^{2}C_{1}\rho _{1}+C_{2}\rho _{2}-C_{3}\rho _{3}, \\
\varsigma &=&\mp \sqrt{\mathcal{E}_{0}^{2}+1}\mathcal{E}_{0}C_{1}\rho
_{1}+C_{2}\rho _{2}+C_{3}\rho _{3},
\end{eqnarray}%
where $C_{k}$ is a constant, the parameters $\rho _{k}$ is%
\begin{eqnarray}
\rho _{1} &=&\exp \frac{\mathcal{E}_{0}^{2}+1}{2\mathcal{E}_{0}}\varkappa ,
\\
\text{ \ }\rho _{2} &=&\exp (\pm \sqrt{\mathcal{E}_{0}^{2}+1}-\frac{\mathcal{%
E}_{0}}{2})\varkappa , \\
\rho _{3} &=&\exp (\mp \sqrt{\mathcal{E}_{0}^{2}+1}-\frac{\mathcal{E}_{0}}{2}%
)\varkappa .
\end{eqnarray}%
Using the expansions obtain 
\begin{equation}
\sqrt{\frac{\int \tilde{\Psi}r^{2}\Psi ds}{\int \tilde{\Psi}\Psi ds}}=\frac{%
\lambda }{2\pi }
\end{equation}

\subsection{Conclusion}

We have calculated the parameter of localization, that is root from the
average radius squared. Surprisingly this parameter in the rotating frame is
few times smaller than that in the lab frame (\ref{rlab}).

\end{document}